\documentclass[12pt]{article}
\renewcommand{\baselinestretch}{1.3}

\newtheorem{theorem}{Theorem}

\newfont{\Mb}{msbm10}
\newcommand{\R}{\mbox{\Mb\symbol{82}}}

\begin{document}

\title{\Large \bf A Note on Segre Types of Second
Order Symmetric Tensors in 5-D Brane-world Cosmology \\ }

\author{
M.J. Rebou\c{c}as\thanks{ {\tt reboucas@cbpf.br}  }, \ \ and \ \
J. Santos\thanks{ {\tt janilo@dfte.ufrn.br} }
\\
\\  \vspace{1mm}
$~^{{\ast}\;}~$Centro Brasileiro de Pesquisas F\'{i}sicas \\
               Rua Dr.\ Xavier Sigaud 150 \\
               22290-180 Rio de Janeiro -- RJ, Brazil \\
\\
$^{\dagger}~$Universidade Federal do Rio Grande do Norte \\
             Departamento de F\'{\i}sica, Caixa Postal 1641 \\
             59072-970 Natal --   RN,  Brazil \vspace{2mm}  \\
        }

\date{\today}

\maketitle

\renewcommand{\baselinestretch}{1.0}
\begin{abstract}
Recent developments in string theory suggest that there might
exist extra spatial dimensions, which are not small nor compact.
The framework of most brane cosmological models is that in which
the matter fields are confined on a brane-world embedded in five 
dimensions (the bulk). Motivated by this we reexamine the
classification of the second order symmetric tensors in $5$--D,
and prove two theorems which collect together some basic results
on the algebraic structure of these tensors in 5-dimensional
space-times. We also briefly indicate how one can obtain, by
induction, the classification of symmetric two-tensors (and the
corresponding canonical forms) on $n$-dimensional ($n > 4$) spaces
from the classification on $4$-dimensional spaces. This is
important in the context of $11$--D supergravity and $10$--D
superstrings.
\end{abstract}

\renewcommand{\baselinestretch}{1.3}

\section{Introduction} 
\label{intro}
\setcounter{equation}{0}

The idea of extra dimensions has a long and honourable history
that goes back to the works of Nordstr\"om, Kaluza and Klein%
~\cite{Nordstrom1914}~--~\cite{Klein1926}. Since that time
Kaluza-Klein-type theories in five or more dimensions have been
used and reemerged over the years in several contexts.

In gauge theories they have been employed in the quest for
unification of the fundamental interactions in physics. In this
case the major idea is that the various interactions in nature
might be unified by enlarging the dimensionality of the
space-time. Kaluza and Klein themselves studied how one could
unify Einstein's theory of gravitation and Maxwell's theory of
electromagnetism in a five-dimensional ($5$--D) framework. The
higher-dimensional Kaluza-Klein approach was later used,
especially among those investigating the eleven-dimensional
($11$--D) supergravity and ten-dimensional ($10$--D) superstrings.

Recent developments in string theory and its extension $M$-theory
have suggested an alternative scenario in which the matter fields
are confined on $3$--D brane-world embedded in $1+3+d$ dimensions
(the bulk). It is not necessary for the $d$ extra spatial
dimensions to be small and compact; and only gravity and other
exotic matter such as the dilaton can propagate in the bulk. This
general picture can be simplified to a $5$--D context where matter
fields are restricted to a $4$--D space-time while gravity acts in
$5$--D~\cite{recent,old}. In this limited $5$--D framework most
work in brane-world cosmology has been done (see, for example,
\cite{BraxBruck2003,Langlois2003} and references therein).

A systematic and independent approach to the $5$--D non-compact
Kaluza-Klein scenario, known as space-time-matter theory (STM),
has also been discussed in a number of papers (see, e.g.,
\cite{WessonLeon1992}~--~\cite{Wesson1999} and references therein;
and also~\cite{PRT}~--~\cite{CRT} ). More recently the equivalence
between STM and brane-world theories has been discussed%
~\cite{Leon2001}, and although they have quite different motivation
for the introduction of extra dimension they are, in a sense,
mathematically equivalent, and solution of STM can be taken over
to $5$-D brane-world theory (see~\cite{Leon2001} for details).

In $4$--D general relativity (GR), it is well known that the
curvature tensor can be uniquely decomposed into three irreducible
parts, namely the Weyl tensor (denoted by $W_{abcd}$), the
traceless Ricci tensor ($S_{ab} \equiv  R_{ab} - \frac{1}{4}\, R\,
g_{ab}$) and the Ricci scalar ($R \equiv R_{ab}\, g^{ab}$). The
algebraic classification of the Weyl part of the Riemann tensor,
known as Petrov classification,  has played a significant role in
the study of various topics in GR. However, for full
classification of curvature tensor of nonvacuum space-times one
also has to consider the Ricci part of the curvature tensor, which
by virtue of Einstein's equations $G_{ab} = \kappa\, {\cal T}_{ab}
+ \Lambda \,g_{ab}\,$ clearly has the same algebraic
classification of both the Einstein tensor $G_{ab} \equiv R_{ab} -
\frac{1}{2}\, R\, g_{ab}\,$ and the energy momentum tensor ${\cal
T}_{ab}$.

The Petrov classification~\cite{petrov} has played an important
role in the investigation of various issues in general
relativity~\cite{ksmh}. The algebraic classification of the Ricci
part $S_{ab}$, known as Segre classification, has been discussed
by several authors~\cite{Hall} and is of great importance in three
contexts. One is in understanding some purely geometrical features
of space-times~\cite{ChurcPlebCorHal}. The second one is in
classifying and interpreting matter field distributions%
~\cite{Hall2Hall1FMMRebTeiRebtei1}. The third is as part of the
procedure for checking whether apparently different space-times
are in fact locally the same up to coordinate transformations%
~\cite{Cartan}~--~\cite{Mac1} (for examples of the use of this
invariant characterization in a class of G\"{o}del-type
space-times~\cite{RT1983} see~\cite{GTappl}).

In 1995 Santos {\em et al.\/}~\cite{SRT1} studied the algebraic
classification of second order symmetric tensors defined on $5$--D
locally Lorentzian manifolds $M$. Their analysis is made from
first principles, i.e., without using the previous classifications
on lower dimensional space-times. However, as concerns the
classification itself their approach is by no means
straightforward.

In a subsequent paper Hall {\em et al.\/}~\cite{HRST1996}
reexamined the algebraic classification of second order symmetric
tensor in $5$--D space-times under different premises. They have
used the knowledge of the algebraic classification of symmetric
two-tensors in $4$--D space-times as a given starting point, and
then shown that the $5$--D algebraic classification obtained
in~\cite{SRT1} could be achieved in a considerably simpler way. A
set of canonical forms for $T_{ab}$ was also presented.
In~\cite{HRST1996} they stated {\em with no proof\/}  a theorem
which collects together a number of important properties of the
algebraic structure of the symmetric two-tensors in $5$--D. The
main goal of the present paper is to offer a detailed proof of
that theorem, and to call the attention to the relevance of the
algebraic classification of an arbitrary second order symmetric
tensor (hereafter denoted by $T_{ab}$) in the context of $5$--D
brane-worlds.  We also briefly indicate how one can obtain, by
induction, the classification of symmetric two-tensors (and the
corresponding canonical forms) on $n$-dimensional ($n > 4$) spaces
from the classification on $4$-dimensional spaces. This is
important in the context of $11$--D supergravity and $10$--D
superstrings.

To make this paper as clear and selfcontained as possible, in the
next section we introduce the notation that will be used, and
present a brief summary of the main results on Segre
classification in $5$--D obtained in our previous articles%
~\cite{SRT1,HRST1996}. These results are required for
section~\ref{mainresults}, where we prove two theorems, related to
the algebraic structure of a second order symmetric two-tensor $T$
defined in a $5$--D locally Lorentzian manifold $M$.

To close this section we note that the results of this paper apply
to any second order real symmetric tensor (such as Einstein, Ricci
and energy momentum) defined on $5$--D locally Lorentzian
manifolds.


\section{Preliminaries}   
\label{pre}
\setcounter{equation}{0}

In this section we shall fix our general setting, define the
notation and briefly review the main results regarding the
classification of a second order symmetric tensor in $5$--D,
required for the last section.

The algebraic classification of symmetric two-tensors $T$ at
a point $p \in M$ can be cast in terms of the eigenvalue problem
\begin{equation}\label{eigeneq}
T^{a}_{\ b}\,V^{b} =\lambda \,\,\delta^{a}_{\ b}\,\,V^{b}\;,
\end{equation}
where $\lambda$ is a scalar, $V^{b}$ is a vector and the mixed
form $T^{a}_{\ b}$  of the tensor $T$ may be thought of as a
linear operator $T: T_{p}(M) \longrightarrow T_{p}(M)$. Throughout
this work $M$ is a real $5$-dimensional space-time manifold
locally endowed with a Lorentzian metric of signature $(- + + +
+), \; T_{p}(M)$ denotes the tangent vectorial space at a point $p
\in M$ and Latin indices range from $0$ to $4$, unless otherwise
stated. Although the matrix $T^a_{\ b}$ is real, the eigenvalues
$\lambda$ and the eigenvectors $V^b$ are often complex. A
mathematical procedure used to classify matrices in such a case is
to reduce them through similarity transformations to canonical
forms over the complex field. Among  the existing canonical forms
the Jordan canonical form (JCF) turns out to be the most
appropriate for a classification of $T^a_{\ b}$.

A  basic result in the theory of JCF is that given an $n$-square
matrix $T$ over the complex field, there exist nonsingular
matrices $X$ such that
\begin{equation}
 \label{simtran}
    X^{-1}T\,X = J \;,
\end{equation}
where $J$, the JCF of $T$, is a block diagonal matrix, each block
being of the form
\begin{equation}   \label{jblock}
 J_{r}(\lambda_{k})=\left[
 \begin{array}{ccccc}
 \lambda_{k} &    1        &   0    & \cdots & 0 \\
      0      & \lambda_{k} &   1    & \cdots & 0 \\
             &             &        & \ddots &   \\
      0      &    0        &   0    & \cdots & 1 \\
      0      &    0        &   0    & \cdots & \lambda_{k}
  \end{array}
  \right]\;.
\end{equation}
Here $r$ is the dimension of the block and $\lambda _{k}$ is the
$k$-th root of the characteristic equation  $\mbox{det}(T -
\lambda I)=0$. Hereafter  $T$ will be the real matrix formed with
the mixed components $T^{a}_{\ b}$ of a second order symmetric
tensor.

A Jordan matrix $J$ is uniquely defined up to the ordering of
the Jordan blocks. Further, regardless of the dimension
of a specific Jordan block there is one and only one
independent eigenvector associated to it.

In the Jordan classification two square matrices are said to be
equivalent if similarity transformations exist such that they can
be brought to the same JCF. The JCF of a matrix gives explicitly
its eigenvalues and makes apparent the dimensions of the Jordan
blocks. However, for many purposes a somehow coarser
classification of a matrix is sufficient. In the Segre
classification, for example, the value of the roots of the
characteristic equation is not relevant --- only the dimension of
the Jordan blocks and the degeneracies of the eigenvalues matter.
The Segre type is a list $\{r_1 r_2 \cdots r_m\}$ of the
dimensions of the Jordan blocks. Equal eigenvalues in distinct
blocks are indicated by enclosing the corresponding digits inside
round brackets. Thus, for example, in the degenerated Segre type
$\{(31)1\}$ four out of the five eigenvalues are equal; there are
three linearly independent eigenvectors, two of which are
associated to Jordan blocks of dimensions 3 and 1, and the last
eigenvector corresponds to a block of dimension 1.

In classifying symmetric tensors in a Lorentzian space two
refinements to the usual Segre notation are often used.
Instead of a digit to denote the dimension of a block with
complex eigenvalue a letter is used, and the digit corresponding
to a timelike eigenvector is separated from the others
by a comma.

In this  work we shall deal with one type of pentad of vectors,
namely  the semi-null pentad basis $\{{\bf l},{\bf m},{\bf x},{\bf
y},{\bf z}\}$, whose non-vanishing inner products are only
\begin{equation}
\label{inerp} l^{a}m_{a} = x^{a}x_{a} = y^{a}y_{a} = z^{a}z_{a} =
1\;.
\end{equation}

At a point $p \in M$ the most general decomposition of $T_{ab}$ in
terms of semi-null pentad basis for symmetric tensors at $p \in M$ is
manifestly given by~\cite{SRT1}
\begin{eqnarray} \label{Tabgen}
T_{ab} & = & \sigma_{1}\,l_a l_b + \sigma_2\,m_a m_b   +
\sigma_3\,x_a x_b + \sigma_{4}\,y_a y_b + \sigma_{5}\,z_a z_b +
2\,\sigma_{6}\,l_{(a}m_{b)}                   \nonumber \\ &  & +
\,2\,\sigma_{7}\,l_{(a}x_{b)} + 2\,\sigma_{8}\,l_{(a}y_{b)} +
 2\,\sigma_{9}\,l_{(a}z_{b)}  + 2\,\sigma_{10}\,m_{(a}x_{b)} +
 2\,\sigma_{11}\,m_{(a}y_{b)}                  \nonumber \\
&  & +\, 2\,\sigma_{12}\,m_{(a}z_{b)} +
2\,\sigma_{13}\,x_{(a}y_{b)} + 2\,\sigma_{14}\,x_{(a}z_{b)}+
2\,\sigma_{15}\,y_{(a}z_{b)} \,, \label{rabgen2}
\end{eqnarray}
where the coefficients $\sigma_1, \ldots ,\sigma_{15} \in \R $.

We can now present the classification of a symmetric two-tensor in
$5$--D by recalling the following theorem, proved in%
~\cite{SRT1,HRST1996}:
\begin{theorem} \label{Stypes}
Let $M$ be a real five-dimensional manifold endowed with a
Lorentzian metric $g$ of signature {\rm(}$ - + + + +${\rm)}. Let
$T^a_{\ b}$ be the mixed form of a second order symmetric tensor
$T$ defined at any point $p \in M$. Then $T^a_{\ b}$ takes one of
the following Segre types: $\{1,1111\}$, $\{2111\}$, $\{311\}$,
$\{z\,\bar{z}\,111\}$, or some degeneracy thereof.
\end{theorem}

For each Segre type of the Theorem~\ref{Stypes} a semi-null basis
with nonvanishing inner products~(\ref{inerp}) can be introduced
at $p \in M$ such that $T_{ab}$ as given by~(\ref{Tabgen}) reduces
to one of the following canonical forms~\cite{SRT1,HRST1996}:

\begin{eqnarray}  \hspace{-0.6cm}
\mbox{\bf Segre type}& &\quad\qquad\qquad\mbox{\bf Canonical
form}\nonumber\\ {\{}1,1111\}  & T_{ab} = &
2\,\rho_1\,l_{(a}m_{b)} + \rho_2\,(l_{a}l_{b} + m_{a}m_{b}) +
\rho_3\,x_{a}x_{b} + \rho_4\,y_{a}y_{b} + \rho_5\,z_{a}z_{b} \;,
\label{rab11111}  \\
{\{}2111\}  & T_{ab} = & 2\,\rho_1\,l_{(a}m_{b)} \pm l_{a}l_{b} +
\rho_3\,x_{a}x_{b} + \rho_4\,y_{a}y_{b} + \rho_5\,z_{a}z_{b}\;,
                                                   \label{rab2111}  \\
{\{}311\}   & T_{ab} = & 2\,\rho_1\,l_{(a}m_{b)} + 2\,l_{(a}x_{b)}
+ \rho_1\,x_{a}x_{b} + \rho_4\,y_{a}y_{b} + \rho_5\,z_{a}z_{b}\;,
                                               \label{rab311} \\
{\{}z\,\bar{z}\,111\}& T_{ab} = & 2\,\rho_1\,l_{(a}m_{b)} +
\rho_2\,(l_{a}l_{b} - m_{a}m_{b}) +\rho_3\,x_{a}x_{b} +
\rho_4\,y_{a}y_{b} + \rho_5\,z_{a}z_{b}\;,
                                                \label{rabzz111}
\end{eqnarray}
and the twenty-two degeneracies thereof. Here $\rho_1, \cdots
,\rho_5 \in \R$ and $\rho_2\, \neq 0$ in (\ref{rabzz111}).

To close this section we recall that the $r$-dimensional ($r \geq
2$) subspaces of $T_{p}(M)$ can be classified according as they
contain more than one, exactly one, or no null independent
vectors, and they are respectively called timelike, null and
spacelike $r$-subspaces of $T_p(M)$. Spacelike, null and timelike
$r$-subspaces contain, respectively, only spacelike vectors, null
and spacelike vectors, or all types of vectors.


\section{Main Results and Concluding Remarks}   
\label{mainresults}
\setcounter{equation}{0}

In this section we will prove {\bf Theorem~4.1} stated in%
~\cite{HRST1996} without a proof. To make the presentation simpler
we have divided it into two theorems: one that deals with
eigenvectors of $T^a_{\ b}$, and another that treats the invariant
subspaces of  $T^a_{\ b}$.

\vspace{2mm}
\begin{theorem} \label{rabtheo1}
Let $M$ be a real 5-dimensional manifold endowed with a Lorentzian
metric $g$ of signature {\rm(}$ - + + + +${\rm)}. Let $T^a_{\ b}$
be the mixed form of a second order symmetric tensor $T$ defined
at a point $p \in M$. Then
\begin{enumerate}
\item
$T^a_{\ b}$ has a timelike eigenvector if and only if it is
diagonable over $\R$ at $p$.
\item
$T^a_{\ b}$ has at least three real orthogonal independent
eigenvectors at $p$, two of which (at least) are spacelike.
\item
$T^a_{\ b}$ has all eigenvalues real at $p$ and is not diagonable
if and only if it has an unique null eigendirection at $p$.
\item
If $T^a_{\ b}$ has two linearly independent null eigenvectors at
$p$ then it is diagonable over $\R$ at $p$ and the corresponding
eigenvalues are real.
\end{enumerate}
\end{theorem}
{\bf Proof}
\begin{enumerate}
\item
In the previous section (see also~\cite{SRT1,HRST1996}) we have
shown that if $T^a_{\ b}$ admits a timelike vector it is
diagonable. Reciprocally, if $T^a_{\ b}$ is a diagonal real matrix
one writes down a general real symmetric matrix $g_{ab}$, uses
that $g_{ac}\, T^c_{\ b}$ is symmetric and obtains $g_{ab} = \;
\mbox{diag}\: (\mu_1,\mu_2,\mu_3,\mu_4,\mu_5)$, where $\mu_1,
\cdots ,\mu_5 \in \R$. As $\mbox{det}\: g_{ab} < 0$, then $\mu_a
\not= 0$ \ ($a = 1, \cdots ,5$). As the metric $g$ has signature
$(- + + + +)$ one and only one $\mu_a$ is negative. The scalar
product on $T_p(M)$ is defined by $g_{ab}$, then the norms of the
eigenvectors of $T^a_{\ b}$ have the same sign as $\mu_a \not= 0$.
Since one $\mu_a < 0$, one of the eigenvectors must be timelike.
\item
{}From (\ref{rab11111})~--~(\ref{rabzz111}) one can easily find the
complete sets of linearly independent eigenvectors and associated
eigenvalues for each non-degenerated Segre type. For each type the
orthogonality is ensured by~(\ref{inerp}). In Table~\ref{eingeing}
we present the eigenvectors and the corresponding eigenvalues for
each Segre type. For all types there are at least three orthogonal
eigenvectors, two of which are spacelike.
\begin{table}
\caption{Segre types, eigenvectors and associated eigenvalues.}
\label{eingeing}
\begin{center}
\begin{tabular}{|c|l|l|} \hline
Segre types & eigenvectors & eigenvalues  \\  \hline \hline
$\{1,1111\}$ & ${\bf l}-{\bf m}$, ${\bf l}+{\bf m}$,
             ${\bf x}$, ${\bf y}$, ${\bf z}$  &  $\rho_1-\rho_2$,
    $\rho_1+\rho_2$, $\rho_3$, $\rho_4$, $\rho_5$ \\ \hline
$\{2111\}$   & ${\bf l}$, ${\bf x}$, ${\bf y}$, ${\bf z}$  &
$\rho_1$, $\rho_3$, $\rho_4$, $\rho_5$   \\  \hline $\{311\}$ &
${\bf l}$, ${\bf y}$, ${\bf z}$ & $\rho_1$, $\rho_4$, $\rho_5$  \\
\hline $\{z\,\bar{z}\,111\}$ & ${\bf l}-\,i\,{\bf m}$, ${\bf
l}+\,i\,{\bf m}$, ${\bf x}$, ${\bf y}$, ${\bf z}$ & $\rho_1-
i\rho_2$, $\rho_1+i\rho_2$, $\rho_3$, $\rho_4$, $\rho_5$
\\  \hline
\end{tabular}
\end{center}
\end{table}
We notice that degeneracies between eigenvalues give rise
to eigenspaces (subspaces containing only eigenvectors),
in which the number of eigendirections increases.
\item
The non-diagonable, non-degenerated Segre types of $T^a_{\ b}$ are
$\{2111\}$ and $\{311\}$. From {\bf (ii)} we learn that both have
only one  null eigenvector ${\bf l}$. Even when there are
degeneracies involving the eigenvalues associated to the null
eigenvector for these types we still have only one null direction
invariant under $T^a_{\ b}$. The converse is clear if one notes
that besides the types $\{2111\}$ and $\{311\}$ the possible types
in which one can have null eigenvectors are the degeneracies of
the type $\{1,1111\}$ involving the eigenvalue associated to the
timelike eigenvector. However, even in the simplest degenerated
type $\{(1,1)111\}$ there exist {\em two} null independent
eigendirections. So, these degenerated types should not be
considered.
\item
Let ${\bf k}$ and ${\bf n}$ be the two linearly independent null
eigenvectors of $T^a_{\ b}$ at $p$ with respective eigenvalues
$\mu$ and $\nu$. The symmetry of $T_{ab}$ implies that $\mu =
\nu$. Moreover,  ${\bf k} + {\bf n}$ and ${\bf k} - {\bf n}$ are
also linearly independent eigenvectors, one of which can easily be
shown to be timelike. Therefore, from {\bf (i)} $T^a_{\ b}$ is
necessarily diagonable.
\end{enumerate}

\begin{theorem} \label{rabtheo2}
Let $M$ be a real 5-dimensional manifold endowed with a Lorentzian
metric $g$ of signature {\rm(}$ - + + + +${\rm)}. Let $T^a_{\ b}$
be the mixed form of a second order symmetric tensor $T$ defined
at a point $p \in M$. Then
\begin{enumerate}
\item
There always exists a 2--D spacelike subspace of $T_p(M)$
invariant under $T^a_{\ b}$.
\item
If a non-null subspace $\cal V$ of $T_p(M)$ is invariant under
$T^a_{\ b}$, then so is its  orthogonal complement ${\cal
V}^{\perp}$.
\item
There always exists a 3--D timelike subspace of $T_p(M)$ invariant
under $T^a_{\ b}$.
\item
$T^a_{\ b}$ admits a $r$-dimensional {\rm(}$r=2, 3,4${\rm )} null
invariant subspace $\cal N$ of $T_p(M)$ if and only if $T^a_{\ b}$
has a null eigenvector, which lies in $\cal N$.
\end{enumerate}
\end{theorem}
{\bf Proof}
\begin{enumerate}
\item
Indeed, the spacelike eigenvectors ${\bf y}$ and ${\bf z}$
referred to in the item {\bf (ii)} of the above theorem
\ref{rabtheo1} span a $2$--D spacelike subspace of $T_p(M)$
invariant under $T^a_{\ b}$.
\item
Let $r$ be the dimension of the non-null subspace $\cal V$, and
${\bf e_{i} }$ ($i=1, \cdots ,r$) a basis of it. Let ${\bf
e^{\perp}_{\mu} }\,$  ($\mu = r+1, \cdots ,5$) be a basis of
${\cal V}^{\perp}$. As $\cal V$ is invariant under $T^a_{\ b}$ the
vectors $T\:{\bf e_{i} } \in {\cal V}$, and therefore ${\bf
e^{\perp}_{\mu} } . T\:{\bf e_{i} } = 0$, which together with the
symmetry of $T_{ab}$ leads to ${\bf e_{i} }.T\:{\bf
e^{\perp}_{\mu} }= 0$. Then for any vector ${\bf v} \in {\cal
V}^{\perp}$ we have $T\:{\bf v} \in {\cal V}^{\perp}$, i.e. ${\cal
V}^{\perp}$ is also invariant under $T^a_{\ b}$.
\item
Indeed, from {\bf (i)} and {\bf (ii)} it follows that the subspace
of $T_p(M)$ spanned by ${\bf l}$, ${\bf m}$ and ${\bf x}$ is
invariant under $T^a_{\ b}$  given in eqs.\ (\ref{rab11111})~--%
~(\ref{rabzz111}).
\item
Clearly every $r$-dimensional {\rm(}$r=2, 3, 4${\rm )} null
subspace of $T_{p}(M)$ can be spanned by a null vector ${\bf n}$
(say) and a set of $r-1$ orthogonal spacelike vectors ${\bf
x_{i}}$ such that ${\bf n} . {\bf x_{i} } = 0 $. Consider the
2-dimensional invariant null subspace generated by such a pair of
basis vectors ${\bf n}$ and ${\bf x}$. Thus
\begin{eqnarray}
T^{a}_{\ b}\, n^b &=& \alpha_1 \, n^{a} + \alpha_2 x^a \,,
\label{2null1} \\ T^{a}_{\ b}\, x^b &=& \beta_1 \,  n^{a} +
\beta_2 x^a  \,, \label{2null2}
\end{eqnarray}
where $\alpha_1\,,\alpha_2\,,\beta_1\,, \beta_2 \in \R$  and ${\bf
n}. {\bf x} ={\bf n}. {\bf n} =0$. Since $T_{ab}$ is symmetric we
have $x_a T^a_{\ b} n^b = n_a T^a_{\ b} x^b$, which together with
(\ref{2null1}) and (\ref{2null2}) yields $\alpha_2 =0$, hence
${\bf n} $ is an eigenvector. The proofs for null subspaces of
dimension 3 and 4 are similar. The converse is easy to show.
Indeed, from eqs.\ (\ref{rab11111})~--~(\ref{rabzz111}) one learns
that the Segre types which admit null eigenvectors are $\{2111\}$,
$\{311\}$ and their degeneracies, and the type $\{(1,1)111\}$ and
its further degeneracies. Moreover, from eqs.\
(\ref{rab11111})~--~ (\ref{rab311}) one finds that for all these
Segre types ${\bf l}$ is an eigenvector and the subspaces of
dimension 2, 3 and 4 spanned by $\{ {\bf l}, {\bf x} \} $, $ \{
{\bf l}, {\bf x}, {\bf y} \} $ and $\{ {\bf l}, {\bf x}, {\bf y},
{\bf z} \} $, respectively, are null and invariant under $T^a_{\
b}$.
\end{enumerate}

Before closing this article we remark that by a similar procedure
to that used in the lemma 3.1 of~\cite{HRST1996}  one can show
that a symmetric two-tensor $T$ defined on an $n$-dimensional ($n
\geq 4$) Lorentz space has at least one real  spacelike
eigenvector. The existence of this eigenvector can be used to
reduce, by induction, the classification of symmetric two-tensors
on $n$-dimensional ($n > 4$) spaces to the classification (and the
corresponding canonical forms) on $4$-dimensional spaces, thus
recovering in a simpler way the Segre types of symmetric
two-tensors on $n$-dimensions and the corresponding set of
canonical forms derived in ref.~\cite{SRT2}. This is important
in the context of $11$--D supergravity and $10$--D superstrings.

Finally we mention that it has sometimes been assumed that the
source term ($T_{ab}$ in $4$--D) restricted to the brane is a
mixture of ordinary matter and a minimally coupled scalar field%
~\cite{HogenColeyHe2002,HoogenIbanez2002}, where the gradient of
the scalar field $\phi_a \equiv \phi_{\,;a}$ is a timelike vector.
In these cases the scalar field can mimic a perfect fluid, i.e.,
it is of Segre type $\{1,(111)\}$. However, according to
ref.~\cite{SRT3}, depending on the character of the gradient of
the scalar field it can also mimic : (i) a null electromagnectic
field and pure radiation (Segre type $\{(211)\}$, when $\phi^a$ is
a null vector); (ii) a tachyon fluid (Segre type $\{(1,11)1\}$,
when $\phi^a$ is a spacelike vector); and clearly (iii) a
$\Lambda$ term (Segre type $\{(1,111)\}$, when the scalar field is
a constant).

\section*{Acknowledgments}
M.J. Rebou\c cas thanks J.A.S. Lima for his kind hospitality at
Physics Department of Federal University of Rio Grande do Norte
(DFTE-UFRN) and A.F.F. Teixeira for reading this manuscript and
indication of some omissions. He also acknowledges CNPq for the
grant under which this work was carried out.


\end{document}